\documentclass{emulateapj}   
\submitted{ApJL, accepted}

\usepackage{rotating}

\def\ltsima{$\; \buildrel < \over \sim \;$}
\def\simlt{\lower.5ex\hbox{\ltsima}} 
\def\gtsima{$\; \buildrel > \over \sim \;$}
\def\simgt{\lower.5ex\hbox{\gtsima}} 
\def\arcsec{\hbox{$^{\prime\prime}$}}
\def\deg{\hbox{$^\circ$}}
\def\cgsflux{erg s$^{-1}$ cm$^{-2}$}

\def\fxfr{$f_{\rm x}$/$f_{\rm r}$}

\def\Chandra{\textit{Chandra}}
\def\XMM{\textit{XMM}}
\def\HST{\textit{HST}}
\def\gb{GB~1428+4217}

\shorttitle{Kiloparsec Scale Jet in $z=4.72$ Quasar GB~1428+4217} 
\shortauthors{Cheung et al.}

\begin{document}

\title{Discovery of a Kiloparsec Scale X-ray/Radio Jet in the $z=4.72$ 
Quasar GB~1428+4217}

\author{
C.~C.~Cheung\altaffilmark{1}, 
\L.~Stawarz\altaffilmark{2,3}, 
A.~Siemiginowska\altaffilmark{4}, 
D.~Gobeille\altaffilmark{5,6}, 
J.~F.~C.~Wardle\altaffilmark{5},
D.~E.~Harris\altaffilmark{4},
D.~A.~Schwartz\altaffilmark{4}
}

\altaffiltext{1}{National Research Council Research Associate, National 
Academy of Sciences, Washington, DC 20001, resident at Naval Research 
Laboratory, Washington, DC 20375, USA; Teddy.Cheung.ctr@nrl.navy.mil}

\altaffiltext{2}{Institute of Space and Astronautical Science, Japan 
Aerospace Exploration Agency, 3-1-1 Yoshinodai, Chuo-ku, Sagamihara, 
Kanagawa 252-5210, Japan}

\altaffiltext{3}{Astronomical Observatory, Jagiellonian University, ul. 
Orla 171, 30-244, Krak\'ow, Poland}

\altaffiltext{4}{Harvard-Smithsonian Center for Astrophysics, 60 Garden 
St., Cambridge, MA 02138, USA}

\altaffiltext{5}{Department of Physics, MS~057, Brandeis University, 
Waltham, MA 02454, USA}

\altaffiltext{6}{Department of Physics, University of South Florida, 
Tampa, FL 33620, USA}

\begin{abstract}

We report the discovery of a one-sided 3.6\arcsec\ (24 kpc, projected) 
long jet in the high-redshift, $z$=4.72, quasar GB~1428+4217 in new 
\Chandra\ X-ray and VLA radio observations. This is the highest redshift 
kiloparsec-scale X-ray/radio jet known. Analysis of archival VLBI 2.3 
and 8.6 GHz data reveal a faint one-sided jet extending out to $\sim$200 
parsecs and aligned to within $\sim$30\deg\ of the \Chandra/VLA 
emission. The 3.6\arcsec\ distant knot is not detected in an archival 
\HST\ image, and its broad-band spectral energy distribution is 
consistent with an origin from inverse Compton scattering of cosmic 
microwave background photons for the X-rays.  Assuming also 
equipartition between the radiating particles and magnetic field, the 
implied jet Lorentz factor is $\approx 5$. This is similar to the other 
two known $z \sim 4$ kpc-scale X-ray jet cases and smaller than 
typically inferred in lower-redshift cases. Although there are still but 
a few such very high-redshift quasar X-ray jets known, for an inverse 
Compton origin, the present data suggest that they are less relativistic 
on large-scales than their lower-redshift counterparts.

\end{abstract}


\section{Introduction\label{section-intro}}

The $z=4.72$ quasar \gb\ (B3~1428+422) was identified in a search for 
high-redshift objects through targeted optical spectroscopy of 
flat-spectrum radio sources \citep{hoo98,fab97}. It is a luminous X-ray 
source with detected X-ray and radio variability characteristic of a 
blazar \citep{fab99,wor06,ver10}. On parsec-scale, VLBA 15 GHz images 
indicate a dominant high brightness temperature, $T_{\rm b} \simeq (4-6) 
\times 10^{11}$ K, core component with a faint one-sided jet-like 
extension \citep[][and references therein]{ver10}. The high luminosity, 
variability, and radio compactness are all properties consistent with 
Doppler beaming of emission from a relativistic jet aligned close to our 
line of sight \citep[see][]{fab99}.

Such high-redshift radio/X-ray sources offer a unique glimpse into 
powerful outbursts from active galactic nuclei (AGN) in the early 
Universe. As the ambient medium into which large-scale jets propagate 
(including host galaxy environments and intergalactic medium) is 
expected to be drastically different at such early epochs 
\citep[e.g.,][]{dey06,mil08}, studies of large-scale jet structures 
allow us to probe radio source interactions with their environment. 
Motivated by X-ray detections of kpc-scale jets in two very 
high-redshift quasars, 1745+624 at $z=3.9$ \citep{che06} and 
GB~1508+5714 at $z=4.3$ \citep{sie03,yua03}, and the noticeable dearth 
of \Chandra\ observations of z$\simgt$2 jet systems 
\citep[e.g.,][]{kat05,har06,mas11}, we began a program to obtain 
arcsecond-resolution radio and X-ray imaging of more such systems 
\citep{che05,che08} with the aim to understand the physics of the 
highest-redshift relativistic jets.

Using \Chandra\ X-ray Observatory and NRAO\footnote{The National Radio 
Astronomy Observatory is operated by Associated Universities, Inc. under 
a cooperative agreement with the National Science Foundation.} Very 
Large Array (VLA) imaging observations of \gb, we discovered an 
X-ray/radio jet separated from the nucleus by 3.6\arcsec\ (24 kpc, 
projected)\footnote{Adopting $H_{\rm 0}=71~$km~s$^{-1}$~Mpc$^{-1}$, 
$\Omega_{\rm M}=0.27$ and $\Omega_{\rm \Lambda}=0.73$, the quasar is at 
a luminosity distance, $D_{\rm L}=44.5$ Gpc and 1\arcsec = 6.59 kpc.}. 
At $z=4.72$, this is the most distant kpc-scale jet imaged in X-rays. No 
significant optical emission is detected from the 3.6\arcsec\ knot in an 
archival {\textit{Hubble Space Telescope (HST)}} image. To probe smaller 
scale emission, we imaged archival very long baseline interferometry 
(VLBI) 2.3 and 8.6 GHz data, revealing a one-sided $\sim$200 parsec long 
jet aligned within $\sim 30\deg$ of the kpc-scale \Chandra/VLA 
structure. In the following, we present these multi-wavelength 
observations (Section~2), and go on to discuss the physical parameters 
of the large-scale outflow in terms of inverse Compton emission models, 
comparing this case to other X-ray detected AGN jets (Section~3).

\section{Radio, X-ray, and Optical Observations}

\begin{figure}
\epsscale{1.0}
\plotone{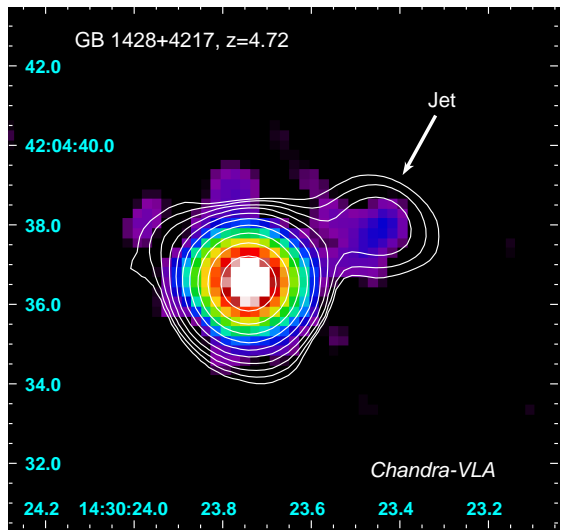}
\figcaption[gb1428xroverlay.ps]{\label{figure-1}
\Chandra\ $0.3-7$ keV (color) and VLA 1.4 GHz (contours) images of \gb\ 
showing the bright core and faint $\sim 3.6\arcsec$ distant jet knot at
$PA \sim 295\deg$. Coordinates are in J2000.0 equinox.
The X-ray data are binned by 1/2 of the native 
0.492\arcsec\ pixels and Gaussian smoothed with kernel radius of 3 
pixels. The 10 radio contours start at 0.17 mJy/bm (4 times the 
off-source rms) increasing by factors of 2 up to 87 mJy/bm (peak is 
155.4 mJy/bm) with circular beamsize = 1.5\arcsec.
}
\end{figure}

As part of a larger survey search for kpc-scale jets in a sample of $z 
\simgt 3.4$ flat-spectrum radio sources \citep{che05}, we obtained VLA 
A-array observations of \gb\ on 2004 Dec 6 (program AC755) at 1.4 and 
4.9 GHz. Data were recorded in two 50 MHz wide channels centered at 
1.385 and 1.465 GHz, and 4.835 and 4.885 GHz, respectively. In the 1.4 
GHz image (2020 s exposure; off-source rms noise = 0.043 mJy/bm) shown 
in Figure~\ref{figure-1}, we discovered a 3.6\arcsec\ distant knot at a 
position angle ($PA$) of $-66\deg$ with a flux density of 1.4 mJy 
(10$\%$ uncertainties are assumed in the radio measurements for the 
knot). Because no counterpart was detected in the shallower 4.9 GHz 
image (710 s), we obtained deeper (1 hr exposure; rms = 0.022 mJy/bm) 
follow-up B-array observations at this frequency on 2008 Jan 13 (program 
S8723) matching the resolution of the A-array 1.4 GHz discovery data. 
The knot is confirmed at 4.9 GHz with a flux density of 0.41 mJy and the 
resultant $1.4-4.9$ GHz radio spectral index is $\alpha_{\rm r} = 1.0 
\pm 0.1$. Model-fitting the radio nucleus and knot in the 1.4 GHz 
($u,v$) data using DIFMAP \citep{she94}, we found that both features are 
unresolved at $\sim 1\arcsec$ resolution and set this as the upper limit 
to the size of the knot.

\begin{table*}
\caption[]{\label{table-1} \gb\ X-ray Spectral Fits}
\begin{center}
\begin{tabular}{lcccc}
\hline \hline
Component Model       &
$\alpha_{\rm x}$      & 
$N_{\rm H}$($z$=4.72) & 
$F$(0.5--2 keV)       & 
$F$(2--10 keV)        \\
(1) & (2) & (3) & (4) & (5) \\
\hline
Core Power-Law             & 0.38$^{+0.09}_{-0.06}$ & $<3.6$ & $5.5 \pm 0.5$  & 16.4$^{+3.3}_{-2.7}$ \\
Core Power-Law with pileup & 0.43$^{+0.25}_{-0.12}$ & $<4.9$ & $6.1 \pm 0.7$  & 16.7$^{+5.8}_{-4.4}$ \\
Jet Power-Law              & 0.7 (fixed)            & --     & $5.7^{+1.8}_{-2.2} \times 10^{-2}$ & $10.8^{+4.2}_{-3.3} \times 10^{-2}$ \\ 
\hline \hline \end{tabular}
\end{center}
(1) Component and source model. All fits assume Galactic $N_{\rm H}$ = 
1.4 $\times 10^{20}$ cm$^{-2}$ fixed \citep{dic90}. Uncertainties are 
90\% for one significant parameter and upper limits are quoted at 
3$\sigma$. \\
(2) We use the definition of spectral index, $\alpha$, as 
$F_{\nu}~\propto~\nu^{-\alpha}$.\\
(3) Intrinsic absorption at $z$=4.72 in units of $10^{22}$ cm$^{-2}$.\\
(4, 5) Unabsorbed flux in the observed energy range in units of 
10$^{-13}$ \cgsflux. The absorbed $F$(0.5--2 keV) values are $14 - 15 
\%$ smaller assuming that the absorber has the $N_{\rm H}$ value set at 
the $3 \sigma$ limit.
\\
\end{table*}

Our \Chandra\ observation of \gb\ was obtained on 2007 Mar 26 
\dataset[ADS/Sa.CXO#obs/07874]({obsid 7874}) as part of a small snapshot 
program targeting four $z>3.6$ flat-spectrum radio quasars with detected 
radio jets in the aforementioned VLA survey \citep{che05}. Of these four 
targets, the X-ray jet detections of \gb\ and one other target 
(PKS~1418--064 at $z=3.7$), were initially reported in \citet{che08}. 
The full results of the survey will be reported elsewhere together with 
the \Chandra/VLA results for an additional seven $z=2-3$ quasars with 
radio jets obtained as part of other programs.

In the 11.7 ks \Chandra\ observation of \gb, we used the nominal 
aim-point of the ACIS-S3 chip and a 1/8th sub-array mode (0.4 s frame 
time) in order to mitigate pileup of the nucleus. We reprocessed the 
data by using {\tt chandra$\_$repro} script in CIAO 4.4 \citep{fru06} 
and assigned the most recent instrument calibration available in CALDB 
4.5. The script also runs {\tt acis$\_$process$\_$events} which applies 
the sub-pixel algorithm and provides the data with the best angular 
resolution required by our analysis. With the correction for 9.4\% 
deadtime, we have 10.6 ks effective exposure time on the source. The 
astrometry was set by adjusting the X-ray core position to that of the 
VLBI radio position from \citet{fey04}, R.A.~=~14$^{\rm h}$30$^{\rm 
m}$23$^{\rm s}$.742, Decl.~=~+42\deg$04'36.49''$ (J2000.0).

\begin{figure}
\epsscale{1.3}
\plotone{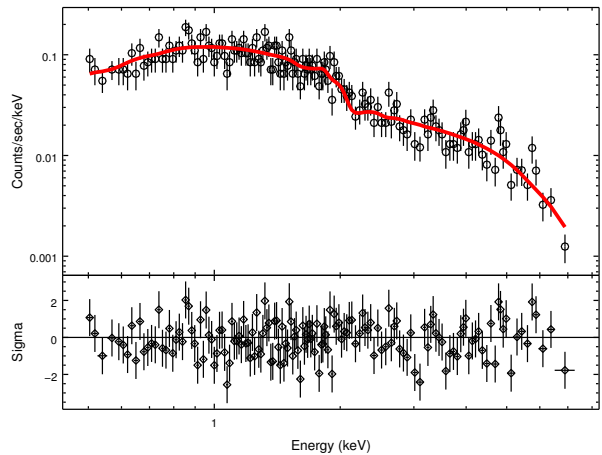}
\vspace{-0.2in}
\figcaption[qso.eps]{\label{figure-2}
\Chandra\ ACIS-S $0.5-7$ keV spectrum of the quasar nucleus in the (top) 
panel. The absorbed power-law model overlaid in red, and the deviations 
from the model in the (bottom) panel. The data were grouped in bins with 
a minimum of 10 counts per bin.
}
\end{figure}

More than 2200 counts were detected from the quasar X-ray core, allowing 
for spectral analysis. Model fitting of the quasar data was performed 
with a 1.5\arcsec\ radius circular aperture. A pie region excluding the 
jet was used to determine the background.  The spectrum was well fit 
with an absorbed power law model (Figure~\ref{figure-2}), with fixed 
Galactic absorption, $N_{\rm H}$. The \Chandra\ derived parameters are 
consistent with the \XMM\ results published by \citet{wor06}, with the 
power-law spectral index slightly smaller in our \Chandra\ spectrum, 
which may be due in part to pileup.  Applying the jdpileup model 
\citep{dav01}, the core is only 2.7$\%$ piled up and the resulting 
parameters are essentially unchanged, but the statistical uncertainties 
increased (see Table~\ref{table-1}).

X-ray emission coincident with the $3.6\arcsec$ distant radio jet knot 
is apparent in the smoothed \Chandra-ACIS image (Figure~\ref{figure-1}). 
The charge transfer readout streak is at $PA = 43\deg$, i.e., almost 
perpendicular to the jet axis, so does not contaminate the knot's 
emission. For the knot, we defined an elliptical aperture region giving 
$20.3 \pm 4.8$ net counts in the $0.3-7$ keV range. The spectrum of this 
emission extends up to a rest frame energy of $\sim 6.5~(1+z)$ keV = 37 
keV, and is quite soft with 14.9 net counts between $0.5-2$ keV and only 
4.4 between $2-7$ keV. The fluxes resulting from a spectral fit assuming 
a single power law with $\alpha_{\rm x}=0.7$ frozen and Galactic $N_{\rm 
H}$ are presented in Table~\ref{table-1}. For the given flux of the jet, 
its isotropic luminosity as measured in the observer rest frame in the 
$0.5 - 10$ keV band would be $4.0 \times 10^{45}$\,erg\,s$^{-1}$.

To help constrain the overall spectral energy distribution (SED) of the 
3.6\arcsec\ knot, we analyzed an archival \HST\ WFPC2 image from 1999 
Feb 8 (total exposure: 9.4 ks) with the F814W filter (program 7266). 
Within a circular aperture ($r=0.5\arcsec$) centered on the knot, we 
found no significant excess optical emission with respect to the 
background measured from six adjacent regions.  From the aperture count 
rates, we derive an aperture corrected \citep{hol95} upper limit of 
$<0.13$ $\mu$Jy ($3 \sigma$) at $3.74 \times 10^{14}$ Hz from the knot.

\begin{figure*}
\epsscale{1.02}
\plotone{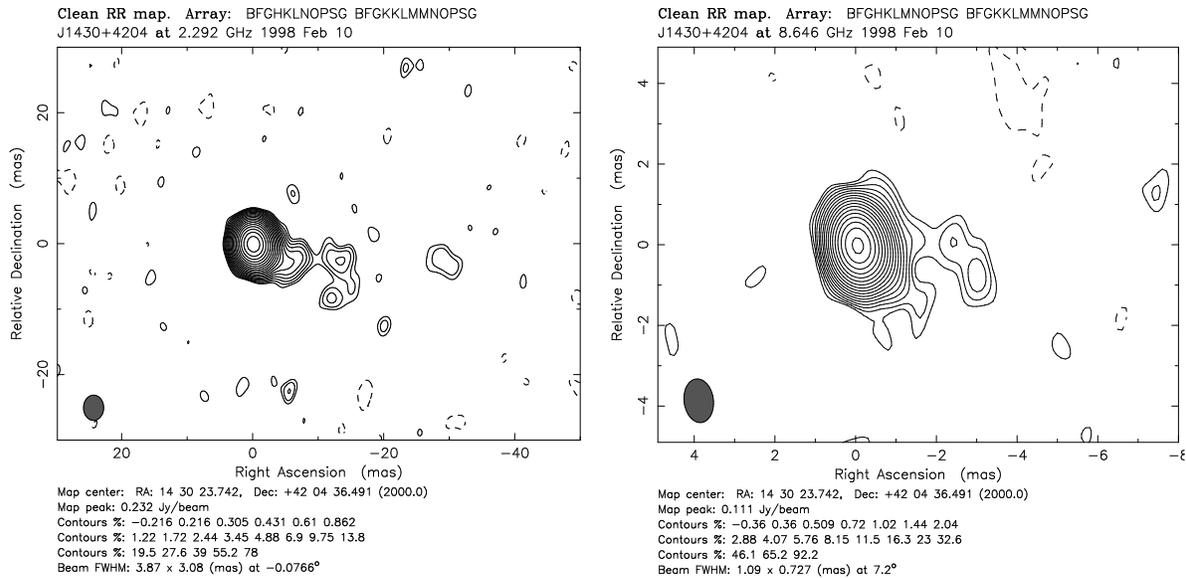}
\figcaption[gb1428vlbi.eps]{\label{figure-3}
VLBI 2.3 GHz (left) and 8.6 GHz (right) images of \gb\ from reprocessing 
USNO data \citep{fey04}. Note the difference in scales. The contours 
begin at 0.5 (3.87 mas $\times$ 3.08 mas) ) and 0.4 mJy/bm (1.09 mas 
$\times$ 0.73 mas), respectively, and increase by factors of $\sqrt{2}$. 
The bright core with the one-sided pc-scale jet to the west is apparent.
}
\vspace{+0.25in}
\end{figure*}

Finally, to search for radio jet emission on smaller scales, we analyzed 
VLBI 2.3 and 8.6 GHz data for \gb\ obtained in 1998 Feb 10 as part of 
the USNO RRFID\footnote{This research has made use of the United States 
Naval Observatory (USNO) Radio Reference Frame Image Database (RRFID).} 
project \citep{fey04}. The calibrated ($u,v$) data were reimaged and 
self-calibrated, revealing a one-sided jet extending west of the core 
(Figure~\ref{figure-3}). The jet is visible in the 2.3 GHz map out to 30 
mas ($\sim$200 pc, projected). This jet can be modeled as three knots at 
$PA = -96\deg$ to $-117\deg$ with a total flux density of 9.5 mJy. 
Utilizing the closest (and brightest) jet knot, we measure a jet to 
counterjet flux ratio defined as the peak / ($3 \times$ rms noise) = 3.3 
mJy / $3 \times 0.16$ mJy $\approx 7$. Most of the pc-scale jet is 
resolved out in the higher resolution 8.6 GHz image, where the jet 
extends out to only 3 mas (2.4 mJy). On smaller scale, \citet{ver10} 
found a faint $\sim 1$ mas extension in VLBA 15 GHz data also in the 
same direction. The pc-scale jet deviates by about $30\deg$ from the 
large-scale \Chandra/VLA knot; however, its existence makes the jet 
interpretation for the kpc-scale emission more likely.

\section{Discussion}

The kpc-scale radio and X-ray jet reported here in the $z=4.72$ blazar 
\gb\ is the highest-redshift example thus far. Together with the blazar 
nature of its core emission (Section~1), the one sidedness of the 
kpc-scale and $\sim$200 pc scale jet seen in our VLBI images imply that 
the jet is relativistic and probably aligned at a small angle, $\theta$, 
to our line of sight. For $\theta \simlt 20\deg$, the 3.6\arcsec\ knot 
distance from the core corresponds to 24/(sin~$\theta$) kpc $\simgt 70$ 
kpc, deprojected for the detected jet. This is far enough from its 
parent host galaxy that the most significant source of seed photons for 
inverse Compton scattering is the cosmic microwave background (CMB). 
This is especially relevant at $z$=4.72, where the CMB energy density 
for an observer at rest at the source redshift is $u^{*}_{\rm CMB} = 4.2 
\times 10^{-13}~(1+z)^{4} = 4.5 \times 10^{-10}$ erg cm$^{-3}$, i.e., 
1070 times greater than it is locally.

The radio and X-ray morphology of \gb\ at $z=4.72$ is nearly identical 
to that of the $z=4.3$ quasar GB~1508+5714 with similar X-ray 
\citep{sie03,yua03} and radio \citep{che04,che05} luminosities in the 
jet. Both show single radio features separated from the bright nucleus 
and display steep spectra (with spectral indices, $\alpha_{\rm r} = 1.0 
\pm 0.1$ and $1.4 \pm 0.2$, respectively). Because of poor statistics, 
the X-ray spectrum in \gb\ is unconstrained and can not be compared with 
that of GB~1508+5714 where $\alpha_{\rm x} \approx 0.9 \pm 0.4$ was 
determined. In \gb, we calculate the ratio \fxfr\ $\equiv (\nu_{\rm x} 
F_{\rm x})$ / $(\nu_{\rm r} F_{\rm r})$ = 205, using monochromatic flux 
densities at 1 keV and 1.4 GHz. This is larger than the $z=4.3$ case 
where \fxfr\ = 158, and larger than found in other lower-$z$ examples 
\citep[cf.,][]{che04,mas11}. Equivalently, \fxfr\ = 205 corresponds to a 
radio to X-ray spectral index, $\alpha_{\rm rx} = 0.72$. Although 
$\alpha_{\rm rx}$ is just consistent with the limit on the radio to 
optical spectral index, $\alpha_{\rm ro} > 0.73$, it is smaller than the 
radio spectral index, $\alpha_{\rm r} = 1.0$, indicating that the X-rays 
are not a simple extension of the radio synchrotron spectrum.

The X-ray emission could correspond to either inverse Compton emission 
of the low-energy electrons involving CMB target photon field 
\citep[`IC/CMB' model;][]{tav00,cel01}, or an additional synchrotron 
component due to a higher-energy electron population 
\citep[e.g.,][]{sta04,hard06}. Overall, the broad-band spectral indices 
appear consistent with the (1+$z$)$^{4}$ amplification of the CMB energy 
density, \fxfr\ $\sim (\delta/\Gamma)^{2}\, u'_{\rm CMB}/u'_{\rm B} 
\propto (1+z)^{4}~(\delta/B)^{2}$ (with the Doppler factor, $\delta$, 
the bulk Lorentz factor, $\Gamma$, and the CMB and magnetic field energy 
densities in the jet rest frame, $u'_{\rm CMB} = \Gamma^{2} \, 
u^{*}_{\rm CMB}$ and $u'_{\rm B} = B^{2}/8\pi$, respectively), as would 
be expected in the IC/CMB model \citep[e.g.,][]{sch02,che04}. Following 
\citet{che04}, we apply the IC/CMB model assuming also equipartition 
with a relativistic electron spectrum with power-law slope, $p = 2 
\alpha + 1 = 3$, extending down to a minimum $\gamma =10$, in a sphere 
with an apparent size of $1.8 \times 1\arcsec$ for the \gb\ jet knot 
\citep[see][]{mar83}. For the case where the jet Lorentz factor is set 
equal to the Doppler factor, we derive $\Gamma = \delta = 4.7$, and a 
magnetic field in the jet rest frame, $B=35$ $\mu$G. These parameters 
are fairly insensitive to the extrapolation of the $p=3$ slope down to 
low energies and our conservative upper limit to the radio knot size.  
Assuming instead a typical observed radio jet spectral index of 0.75 at 
lower-redshifts \citep[following][]{kat05}, and the knot diameter is 
$0.6\times$ smaller than the assumed 1\arcsec\ limit, we obtain $B=21$ 
$\mu$G and $\delta = 6.3$. These estimates assume a single X-ray/radio 
emitting zone applies to our $\sim 1\arcsec$ (6.6 kpc) angular 
resolution element.

Similarly applying the IC/CMB model to the other high-redshift X-ray 
jet, GB~1508+4714 \citep[$z=4.3$;][]{che04} and 1745+624 
\citep[$z=3.9$;][]{che06}, they derived $\delta \sim 3-5$. These values 
are smaller than the typical ones ($\delta \sim 5-10$) determined for 
lower-redshift jets \citep[e.g.,][]{kat05}. Although the number of 
quasar X-ray jets found at such very high-redshifts ($z \sim 4 - 5$) is 
still small, the slower large-scale jets implied in our analysis for the 
current sample seem to indicate that they are either: (1) intrinsically 
less relativistic, or (2) decelerate more rapidly out to $\sim$ 10's $-$ 
100's kpc-scales than their lower-redshift counterparts. In the first 
case, this would be consistent with the recent finding that the relative 
abundance of high-power blazars relative to the parent population of 
radio-loud quasars and radio galaxies decreased substantially above $z 
\sim 3$, likely as a result of a decrease of the average bulk Lorentz 
factor of blazar jets in the early Universe \citep[see][]{vol11}. The 
latter possibility could be supported by the fact that the environments 
of high-redshift ($z > 2$) radio sources are believed to reside in more 
inhomogeneous and multi-phase environments than in nearby radio sources 
\citep[see e.g.,][]{ree89,dey06}, and this may manifest in the slower 
large-scale jets due to an enhanced entrainment of the ambient gas when 
the jet propagates through the host galaxy.

An argument put forth against the IC/CMB model comes from observations 
of the unresolved (at \Chandra\ resolution) quasar ``cores'' at 
high-redshift.  In this interpretation for the kpc-scale emission, one 
may expect that the X-ray emission in these cores could also be 
enhanced, as they contain an unresolved portion of the jet (which we 
know are highly relativistic from, e.g., VLBI superluminal motion 
studies).  This could manifest in different X-ray core properties at 
high-$z$ (X-ray, and optical-to-X-ray spectra), but \Chandra\ studies 
thus far show no such trends with redshift \citep[e.g.,][]{lop06}. 
However, these analyses are hampered by the fact that most of the 
sources analyzed so far in this context are only moderately radio-loud, 
and in general, only a small fraction of jets in radio-loud quasars 
extend to kpc-scales \citep[e.g.,][]{bri84,liu02}.

The radio structures of higher-redshift samples are now being 
investigated in more detail \citep[][Gobeille et al., in 
preparation]{gob11}, and will pave the way for further \Chandra\ X-ray 
studies. X-ray observations of a larger sample of kpc-scale radio jets 
over a broad redshift range ($z>2$) should provide further elucidation 
of the correct X-ray emission mechanism which will give us insight on 
the deeper physics contained in the data.  Such high redshift systems 
allow us to study jets which result from the ``earliest'' actively 
accreting black hole systems. As most detections are currently at 
$z\simlt 2$, we have begun to study more distant examples and these 
results will be presented in a future paper.

\acknowledgments
\begin{center}Acknowledgments\end{center}

This research was supported in part by NASA through contract NAS8-39073 
(A.~S., D.~E.~H., D.~A.~S.) and \Chandra\ Award number GO7-8114 issued 
to NRAO and Eureka Scientific (C.~C.~C.) and Brandeis University (D.~G., 
J.~F.~C.~W.) by the \Chandra\ X-Ray Observatory Center, which is 
operated by the Smithsonian Astrophysical Observatory for and on behalf 
of NASA under contract NAS8-39073. Radio astronomy at Brandeis 
University is supported by NSF grant NSF1009261. This work began while 
C.~C.~C.~was supported by a Jansky fellowship from NRAO; his work at NRL 
is supported in part by NASA DPR S-15633-Y.

{\it Facilities:} \facility{CXO ()}, \facility{VLA ()}, \facility{HST 
(WFPC2)}

{}

\end{document}